\documentclass[11pt,showpacs,showkeys,amsmath,amssymb]{revtex4}
\usepackage[usenames,dvipsnames,svgnames,table]{xcolor}
\usepackage{latexsym}
\usepackage{graphicx}
\usepackage{subfigure}

\begin{document}

\title{A model for superconducting sulfur hydrides.}

\author{R. M. M\'{e}ndez-Moreno}
\email{rmmv@ciencias.unam.mx}

\affiliation{Departamento de F\'{\i}sica, Facultad de Ciencias,\\
         Universidad Nacional Aut\'{o}noma de M\'{e}xico,\\
         Avenida Universidad 3000, 04510 M\'{e}xico, D.~F., M\'{e}xico
        }

\begin{abstract}

Due to the value of the isotope shift in sulfur hydrides, a phonon-mediated pairing scenario of superconductivity  is generally accepted for these high-temperature superconductors which is consistent with  the Bardeen-Cooper-Schrieffer (BCS) framework. Knowing that a large electronic density of states enhances $T_c$, generalized Fermi surface topologies are used to increase it. A multi-band  model within the BCS framework is proposed in this work for the description of sulfur hydride superconductors. This model is used to describe some properties of the $H_3S$ superconductor. Strong coupling effects are taken into account with the effective McMillan approximation and the  isotope coefficient is evaluated as function of the coupling parameter as well as other relevant parameters of the model.

\end{abstract}


\maketitle

\section{Introduction}
\vspace{0.2cm}

The research on the possibility of room temperature superconductivity has been invigorated by the discovery of superconducting sulfur hydrides at very high pressure, of about $155GPa$ \cite{Drozdov:2015, Drozdov:2014}. As hydrogen has the smallest atomic mass, it was predicted that metallic hydrogen or hydrogen-rich compounds would be high-Tc superconductors \cite{Ashcroft:2004,Ginzburg:1969,Ashcroft:1968}. Light hydrogen atoms provide the  high frequency phonon modes and a strong electron-phonon coupling. 
It has been shown that $H_2S$ is stable bellow $43 GPa$  and at elevated pressures it decomposes into $H_3S$ and $S$ \cite{Errea:2015, Einaga:2015}. $H_3S$ has been found stable at least up to $300 GPa$ \cite{Duan:2015}. The superconducting temperature of $H_3S$ at pressures above $150 GPa$ is as high as $203K$. This superconductor has the highest transition temperature obtained in high pressure experiments reported until now \cite{Durajski:18}. 

The high value of the isotope exponent of this material evidences a phonon-mediated pairing mechanism, consistent with the BCS framework.
The fact that the critical temperature varies with the isotopic mass was an evidence for the interaction between the electrons and the lattice vibrations \cite{Duan:2014, Durajsky:2015, Yinwei:2014}. These results are a proof that the electron-phonon interaction is an important pairing mechanism in the sulfur hydrides \cite{Gor'kov:2016, Kudryashova:2017,Durajsky:2016b}. The observation that the superconductivity in these superconductors shows a strong sensitivity to the crystal lattice suggest the possibility of unconventional electron-phonon coupling. Phonon properties in  sulfur hydride systems are actually responsible for changes in their properties \cite{Durajsky:2016}.

Understanding of the electronic structure at the Fermi level can  give some useful clues to unravel the fundamental ingredients responsible for the high transition temperature. However, up to now the underlying physical process remains unknown. In this context, it seems crucial to study new ideas that use simplified schematic models to isolate the mechanism(s) that generate high transition temperatures. It had been known that a large value of the density of states at the Fermi level increases the critical temperature value. That is, $T_c$ can be enhanced when the Fermi level is at or close to the energy of a singularity in the density of states, as a van Hove one, which provides a peak in the density of states \cite{Sano:2016,Szczesniaky:2018,Quan:2016}. 

First principles calculations based on the supercell method have been reported, where the authors study the effect of dopping $H_3S$ on superconductivity \cite{Nakanishi:2018}. They show that superconductivity in $H_3S$ can be enhanced by hole doping.
Hirsch and Scalapino signaled the possibility of enhancing the superconducting critical temperature with a two-dimensional structure when the Fermi level was near a singularity \cite{Hirsch:1986}.  A schematic model with generalized Fermi surface topologies, via band overlapping, have also been proposed by us in the weak and intermediate coupling case as a way of increasing the density of states at the Fermi energy. This model is based on the idea that the $T_c$ is enhanced when the Fermi level lies at or close to the energy of a singularity in the density of states. When combined with the Cooper pair equation this model has been shown to account for the higher $T_c$ values obtained with cuprate superconductors \cite{Moreno:1996}. The schematic model model can be taken as a simple device to model the singular behaviour in the density of states.

Pairing symmetry is an important element toward understanding the mechanism of high-$T_c$ superconductivity. Calculations with two and three-dimensional models using the BCS formalism and order parameter with $s$-wave symmetry have been reported for superconducting $H_3S$  \cite{Souza:2016}. It was suggested that the search for better superconductors should be on three dimensional systems in order that the thermal fluctuations be less likely to reduce the  observed $T_c$ \cite{Talantsev:2017}.

The Eliashberg equations are extensions for strong coupling of the original BCS theory. Calculations with this theory can be found where they study the high-$T_c$ in sulfur hydride as a result of the variability in the density of states in the band \cite{Kudryashovb:2017}. Many theoretical calculations based on Eliashberg theory for $H_3S$, can be found in the literature among them in \cite{Capitani:2017}, where the authors also provide an optical spectroscopy study for this material and found spectroscopy evidence that the superconducting mechanism in $H_3S$ is the electron-phonon interaction. It has been shown that $H_3S$ is a very highly optimized electron-phonon superconductor  \cite{Nicol:2015}. As it is known, McMillan numerically solved the Eliashberg non-linear equations at finite temperature in order to obtain the critical temperature for strong coupled superconductors. McMillan got a parametrization which relates the critical temperature to a small number of parameters \cite{McMillan:1968}. This approach, which  is valid for values of the electron-phonon coupling constant $\lambda < 1.5$, was later modified by Allen and Dynes to include values in the strong coupling region up to $\lambda \simeq 2$ \cite{Dynes:1972,Allen:1975,Carbotte:1990}.

Based on electronic band structure obtained for this materials \cite{Duan:2014,Bernstein:2015}, generalized Fermi surface topologies modeled with band overlapping are used in this work as a model of study  for sulfur hydrides. A  multi-band model within the BCS framework is proposed, this proposal, which can be taken as a minimal singularity in the density of states and the BCS framework can lead to higher $T_c$ values, as the energy band overlapping increases the DOS near the Fermi level. We suppose the pairing mechanism is via boson exchange like phonons by example. For physical consistency, an important requirement of the proposed model is that the band overlapping parameter do not be larger than the phonon energy, $E_{ph}$. The model with generalized Fermi surface topologies is now extended to intermediate and strong coupling  with the use of the McMillan effective approach to the Eliashberg equations. In this work we use a momentum independent pairing interaction  supposing that the superconducting order parameter has s-wave symmetry. The model here proposed will be used to describe some properties of sulfur hydride  superconductor $H_3S$, as the isotope mass exponent $\alpha_I$, in terms of the coupling constant and the parameters of the model.

\vspace{0.6cm}
 \section{The model}
 \vspace{0.2cm}

We begin with the gap equation
  \begin{equation}  \label{eq:aa}
    \Delta(k'))= {\sum_k} V(k,k') \Delta(k)\frac{\tanh( E_k/2 k_B T )}{2
    E_k} ,
  \end{equation}
with $V(k,k')$ the pairing interaction, $k_B$ is the Boltzman constant, and $E^2_k =
\epsilon^2_k + \Delta^2_k$, where $\epsilon_k = \hbar^2 k^2/ 2 m$
are the self-consistent single-particle energies.

Then for the electron-phonon interaction, we have considered $V(k,k{\prime})= V_0 $, with $V_0$ a constant, when $|\epsilon_k|$ and $|\epsilon_{k^{\prime}}|~ \leq E_D ~=~k_B
T_D$ and $0$ elsewhere. As usual the attractive BCS interaction is nonzero only for unoccupied orbitals in the neighborhood of the Fermi level $E_F$.  The superconducting order parameter, $\Delta(k) = \Delta(T)$ if $|\epsilon_k| \leq E_{ph}$ and $0$ elsewhere.

With these considerations we propose a model with two overlapping bands. The generalized Fermi sea proposed consists of concentric spheric shells separating occupied orbitals. As a particular distribution in momentum space the following form for has
been considered
\begin{equation}   \label{eq:ab}
      n_k =\Theta(\gamma k_F - k) + \Theta(\gamma k_F
     - k) \Theta(k - \beta k_F ),
\end{equation}
with $k_F$ the Fermi momentum and $0 <  \beta <   \gamma <  1$.  In
order to keep the average number of electron states constant, the
parameters are related in the  system by the equation
 $ 2 \gamma^2 - \beta^2 = 1$, then only one of the relevant parameters
is independent. The distribution in momentum induces one in energy,
$E_{\beta}  < E_{\gamma}$  where  $E_{\beta} = \beta^2 E_F$ and
$E_{\gamma} = \gamma^2 E_F$ . We consider a high frequency electron-phonon 
coupling  $E_{ph}$, with optical phonons and require that the band overlapping be of the order or smaller than the cutoff energy ($E_{ph}$). That means $(1 - \gamma^2) E_F \leq E_{ph}$.  The minimum $\gamma^2$ value consistent
with our model is $\gamma^2_{ph}~ =~1 - E_{ph} / E_F$. The last equation can be written as
$1 - \gamma^2 = 2 \nu \delta$, where $2 \nu  = E_{ph} / E_F$ and $\delta $ is in the range $0< \delta < 1$. While ~ $E_F - E_{\gamma} \leq E_{ph}$,
implies that the energy difference between the anomalously occupied
states must be provided by the material itself. The shallow second band in this model can account for the other experimentally detected bands.

In the last framework the summation in Eq.(\ref{eq:aa}) is
changed to an integration which is done over the ({\it symmetric})
generalized Fermi surface defined above. One gets

\begin{equation}    \label{eq:bb}
\begin{split}
 1~ = ~ \frac{ V_0 D(E)}{2} ~\int_{E_\gamma-
E_{ph}}^{E_\gamma + E_{ph}}~ \tanh
\left(\frac {\sqrt{~\Xi_k}}{2 \it k_B T }  \right)\frac{
d\epsilon_k} { \sqrt{ \Xi_k}}
+ \\
\frac{  V_0 D(E)}{2} ~\int_{E_\beta}^{ E_F} ~ \tanh
\left(\frac {\sqrt{~\Xi_k}}{2 \it
k_B T }  \right) \frac{ d\epsilon_k}{ \sqrt { \Xi_k}}.
\end{split}
\end{equation}

In this equation $\Xi_k = (\epsilon_k  - E_F)^2 + \Delta(T)^2$,
$V_0$ is the effective attractive interaction of the BCS model
and $D(E)$ is the electronic density of states, which will be taken as a constant in the integration range.

 The two integrals correspond to the bands
proposed by Eq. (\ref{eq:ab}).
The integration over the surface at $E_{\gamma}$ in the first band,
is restricted to states in the interval $E_{\gamma} - E_{ph} \leq E_k
\leq {E_{\gamma}+ E_{ph}}$. In the second band, in order to conserve
the particle number, the integration is restricted to the interval
$E_{\beta} \leq E_k \leq {E_F}$, if ~$E_{\gamma}+ E_{ph}>E_F$, with
$E_{\beta}  ~=~ (2~\gamma^2 ~-~1 ) E_F$, in terms of the parameter
$\gamma$ in our model.

The critical temperature is introduced via the Eq. (\ref{eq:bb}) at
$T = T_c$, where the gap becomes $\Delta(T_c) = 0$. At this
temperature Eq. (\ref{eq:bb}) is reduced to
\begin{equation}    \label{eq:cc}
\begin{split}
 1~=~\frac{ V_0 D(E)}{2}~ \int_{E_\gamma - E_{ph}}^{E_\gamma +
E_{ph}} \tanh \left( \frac { \epsilon_k - E_F}{2 \it
k_B T_c} \right)\frac{ d \epsilon_k} {  \epsilon_k  - E_F}+ \\
\frac{ V_0 D(E)}{2}~\int_{E_\beta}^{E_F} \tanh \left(\frac {
\epsilon_k - E_F}{2 \it k_B T_c}\right)\frac{ d \epsilon_k}  {
\epsilon_k  - E_F},
\end{split}
\end{equation}
which is to be numerically evaluated. $D(E)$ is the fermionic density of states and $V_0 $ is the effective attractive interaction of the BCS model. Taking into account that $V_0 D(E) = (\lambda - \mu*) / (1 + \lambda )$, where $\mu^*$ is the Coulomb pseudopotential, the last equation relates $T_c$
to the coupling constant $\lambda$ and to the anomalous occupancy
parameter $\gamma^2$. This relationship determines the $\gamma^2$
values which reproduces the critical temperature of $H_3S$ in the intermediate and strong  coupling region.

At $T = 0K$, Eq.~(\ref{eq:bb}) will be evaluated and $\lambda$
values consistent with the model which reproduce the values obtained for the zero temperature superconducting gap will be obtained:
\begin{equation}  	\label{eq:ee}
\begin{split}
 1  =  \frac{V_0 D(E)}{2}  \sinh^{-1}~ \frac{E_D~+~(\gamma^2 - 1)E_F}{\Delta_0}\\ 
        +  \frac{V_0 D(E)}{2} ~\sinh^{-1}~ \frac{E_D +~(1~-~\gamma^2)E_F} {\Delta_0}\\
        +  \frac{V_0 D(E)}{2} ~ \sinh^{-1}~\frac{2~(1~-~\gamma^2)E_F} {\Delta_0}.
\end{split}       
\end{equation}

The Cooper pair equation and the two band model in this work, are used in order to obtain the isotope mass exponent. The McMillan approach which is valid for values of the coupling constant  $\lambda < 1.5$ is used.  because  results in the literature are quite different \cite{Harshman:(2017),Harshman:(2017b),Hirsch:2015}, the isotope exponent $\alpha$, in the harmonic approximation is also evaluated. The partial isotope coefficient in the harmonic approximation is given by
\begin{equation}
	       \alpha_i = -\frac{d lnT_c}{d lnM_i}
\end{equation}

The equation obtained is
\begin{equation}
  \alpha = 0.5 \left[ 1 + \delta \frac{2 Y - 1 + \frac{\delta + 2 Y - 1}{D}} 
  {\delta(1 - 2 Y) -1 - D} \right], 	
\end{equation}
with $\delta = (1 - \gamma^2) E_{phonon}/E_F$. Where  
\begin{equation}
	D = \sqrt{\delta^2 + 2 ( 2 Y - 1) \delta + 1},
\end{equation}
and the  $Y$ factor is given by
\begin{equation}
	Y = exp[\frac{1.04(1 + \lambda)}{\lambda - \mu^* (1 + 0.62 \lambda)}]
\end{equation}
This equation allows to study the effect of intermediate and strong coupling up to $\lambda =1.5$ in the energy band overlapping model of this work.

The model presented in this section can be useful to describe sulfur hydride superconductors. Ranges for the coupling parameter $\lambda$ in the  intermediate and strong coupling region, $0.5 < \lambda < 1.5$, are taken. Different values of $\mu*$ taken from the literature are used in the calculations. The overlapping parameter $\gamma^2$, values, consistent with the model, are obtained for the material. The relationship between the characteristic parameters will be obtained for $H_3S$ at several pressure (or $T_c$) values. The pairing in the superconducting state is taken as due principally by high-frequency optical modes.

 \vspace{0.6cm}
 \section{Results and discussion}
 \vspace{0.2cm}
 
Next the results of this model are shown,  values of $\mu*$ found in references \cite{Duan:2014,Errea:2015} are taken. The Fermi and Debye energy data  from references \cite{Durajsky:2016b,Kudryashova:2017} were assumed to hold.

\begin{figure*}
\begin{subfigure} 
 \centering
 \includegraphics[width=.4\linewidth]{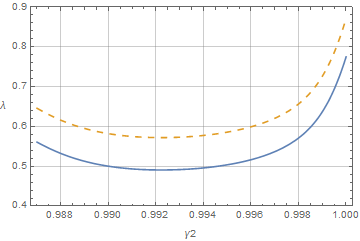}
\end{subfigure} 
\begin{subfigure} 
 \centering
 \includegraphics[width=.4\linewidth]{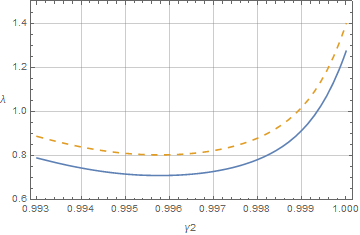}
\end{subfigure} 
\caption{Values of $\lambda$ obtained with the model in this work obtained via the  $Tc$ curves in terms of the parameter $\gamma^2$. In figure(a), $E_F = 16eV$ and $E_{phonon} = 0.214 eV$. In figure(b) $E_F$ is the same as in (a), and  $E_{phonon} = 0.107 eV$}
\end{figure*}

In Figs. 1 the behavior of the coupling parameter $\lambda$ as a function of the band overlapping $\gamma^2$ are shown for the  $H_3S$ at $150 Gpa$, where the critical temperature is about $203 K$. The range  of  $\lambda$ values obtained with our model are in the intermediate and strong coupling region. The values $\mu^* = 0.1,~ 0.16$ were taken in both figures, the full curves are for the small value of $\mu^*$. In 
Fig. (a) the Fermi energy is taken as 16 eV and the phonon energy is 0.214 eV as reported in references \cite{Durajsky:2016b,Akashi:2015}. In Fig. (b) the phonon energy is half the value and $E_F$ is the same as in Fig. (a). In the last figure the required values of the coupling constant are larger than in Fig. (a) but in the region of the Mc-Millan approximation. 

\begin{figure}[t]
\begin{subfigure} 
 \centering
 \includegraphics[width=.4\linewidth]{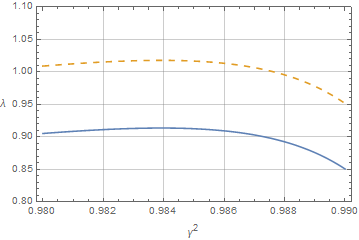}
\end{subfigure} 
\begin{subfigure} 
 \centering
 \includegraphics[width=.4\linewidth]{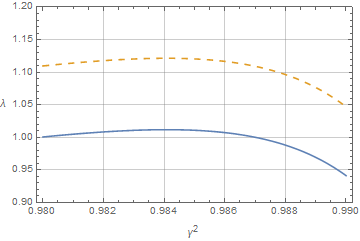}
\end{subfigure} 
\caption{ The curves for $\Delta_0$ are shown for two values of the $\mu^*$. In figure(a), $\Delta_0 = 45$ as in reference and $\Delta_0 = 60$. } \vspace*{1cm}
\end{figure}
In Figs. 2 the zero temperature gap, $\Delta_0$, is shown as function of the coupling constant. The values taken for  $E_F$ and $E_{phonon}$ in both Figs. 2 are the same as in Fig. 1(a). Two values of $\mu*$ were taken, the full curves are for $\mu* = 0.1$ and $\mu* = 0.16$ for the dashed ones. In Fig. (a) $\Delta_0 = 45$ while in Fig. (b) $\Delta_0 = 60$. The larger value of the gap in Fig. 2(b) requires  bigger coupling constant values, but consistent with the  values of Fig. 1(a).

\begin{figure}
\setlength{\unitlength}{0.4cm}
\centering
 \includegraphics[width=.4\linewidth]{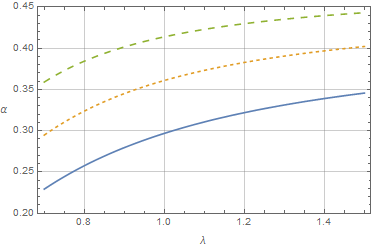}

 \caption{ The isotope mass exponent as function of the $\lambda$, using $\delta$ values ($\gamma^2$) obtained with our model and $E_{phonon}/E_F$ taken from the literature.} \vspace*{1cm}
\end{figure}

In Fig. 3 The isotope mass exponent is evaluated as function of the coupling constant for different values of the ratio $E_{phonon}/E_F$ found in the literature and the overlapping parameter $\gamma^2$, that is the parameter $\delta = (1 - \gamma^2) E_{phonon}/E_F$. The curves for $\delta = 0.01, 0.02. 0.04$ are shown from up to down, $\mu^* = 0.1$ in all the curves.  With $\lambda > 1$ we get $0.3 < \alpha < 0.45$, this values are smaller but close to the BCS result $\alpha = 0.5$.  The mass exponent is greater for  greater values of the ratio $E_{phonon}/E_F$. Our results agree with those obtained using other models in the literature. The values obtained in this work are in the range of values obtained in reference \cite{Akashi:2015}, at different pressures. The results  and $\alpha = 0.25$ of reference \cite{Gorkov:2018} are also in the range of values we obtained for the mass exponent.

In order to increase the density of states at the Fermi level for high-temperature superconductors, we presented a model with generalized Fermi surface topologies  obtained via band-overlapping. The order parameter is supposed to have s-wave symmetry. We use this model to describe sulfur hydride superconductors, within the BCS framework and the model is extended to include intermediate coupling with the McMillan approximation. The model we have used has two overlapping bands at the Fermi level  The behavior of the coupling parameter $\lambda$  as function of the overlapping parameter $\gamma^2 $ is reported, for different  samples. The $\lambda$ values consistent with the model are in the intermediate coupling region. A non-standard electron-phonon coupling  is considered and the minimum band overlapping parameter is consistent with this value. The band overlapping allows the improvement of the results obtained with a s-wave mean-field approximation, in a scheme in which the electron-phonon interaction is the relevant one for high-$T_c$  mechanism {\it i.e.}, the energy scale of the band overlapping
$(1 - \gamma^2)E_F$, is of the order of the phonons energy. This energy is then the overall scale that determines the highest $T_c$ and gives consistency to the model because it requires an energy scale accessible to the lattice. The $H_3S$ isotope exponent  evaluated  was compared with other calculations and with experimental values reported. We were not able to get a larger value of the isotope mass exponent with this generic model. That suggest that it is necessary to modify the model in order to take into account additional elements. However this schematic model can be a guide for a more detailed study of these materials with, for example, Eliashberg theory.

\end{document}